\documentclass[aps,preprint]{revtex4}
\usepackage{epsfig}
\usepackage{amsmath}
\usepackage{graphicx}
\usepackage{dcolumn}
\usepackage{natbib}
\usepackage{amssymb}
\begin{document}

\newcommand{\calCfrac}{{{\cal C}_{\mathrm{frac}}}}
\newcommand{\zetaPlus}{{\zeta_{+}}}
\newcommand{\zetaMinus}{{\zeta_{-}}}
\newcommand{\zetaKPZ}{{\zeta_{KPZ}}}
\newcommand{\zetaEW}{{\zeta_{EW}}}

\title{Crossover behavior in interface depinning}

\author{Y. J. Chen}
\affiliation{LASSP, Physics Department, Cornell University, Ithaca, NY 14853-2501, United States}

\author{Stefano Zapperi}
\affiliation{Center for Complexity and Biosystems,
Department of Physics, University of Milano, via Celoria 26, 20133 Milano, Italy}
\affiliation{CNR - Consiglio Nazionale delle Ricerche, IENI, Via R. Cozzi 53, 20125,
Milano, Italy}
\affiliation{ISI Foundation, Via Alassio 11/c 10126 Torino, Italy}
\affiliation{Department of Applied Physics, Aalto University,
P.O. Box 14100, FIN-00076, Aalto, Finland}

\author{James P. Sethna}
\affiliation{LASSP, Physics Department, Cornell University, Ithaca, NY 14853-2501, United States.}

\begin{abstract}
We study the crossover scaling behavior of the height-height correlation function in interface
depinning in random media. We analyze experimental data from a 
fracture experiment and simulate an elastic line model with non-linear couplings and disorder.
Both exhibit a crossover between two different universality classes. 
For the experiment, we fit a functional form to the universal crossover scaling function. For the model,
we vary the system size and the strength of the non-linear term,
and describe the crossover between the two universality classes
with a multiparameter scaling function. Our method provides a general strategy to
extract scaling properties in depinning systems exhibiting crossover phenomena.
\end{abstract}

\maketitle
\section{Introduction}
\label{sec:Intro}

Driven interfaces in random media display intriguing scaling laws that are common to a 
wide variety of phenomena, including fluid imbibition, crack front roughening, dislocation hardening, superconducting flux lines, the equilibrium motion of piles of rice down an incline, and domain wall motion in magnets~\cite{barabasi95fractal,kardar97interfaces}.  The scaling laws are commonly associated with an
underlying depinning critical point  that has been elucidated by simple models for interface dynamics.
These models  have been extensively studied using continuum simulations~\cite{leschhorn97,rosso01,rosso02,rosso03}, cellular automata~\cite{tang92,leschhorn93,leschhorn92,leschhorn94,leschhorn96,leschhorn97}, 
and field-theoretic $\epsilon$ expansions~\cite{narayan92,narayan93,narayan94,leschhorn97,chauve00,chauve01,ledoussal02},
providing a sophisticated picture of the non-equilibrium phase transition and of the 
different universality classes. 

The interface morphology is usually characterized by the roughness exponent $\zeta$, resulting from a coarse graining operation of the interface height function $h(x)$. Namely, when we change all length scales by a factor $b$, or $x \rightarrow bx$, then statistically $h \rightarrow b^{\zeta} h$ -- hence
\begin{equation}
h(x) \sim b^{-\zeta} h(bx).
\end{equation}
For many experiments and simulations, it is convenient to measure $\zeta$ by computing the height-height correlation of the interface 
\begin{equation}
C(r) = \langle(h(x+r)- h(x))^2 \rangle \sim r^{2 \zeta}
\label{eq:heightzeta}
\end{equation}
(In Sec.~II~\cite{SantucciGHTSM14} we shall study a system with
{\em anomalous scaling}, where the power law exhibited by $C(r)$ differs
from the universal rescaling exponent $\zeta$.
Rather than rescaling $h$ in such systems, one studies the rescaling of the
correlation function directly,
$C(r) \sim b^{-2\zeta} C(br) \sim r^{2 \zeta}$.
These systems are multi-affine~\cite{SantucciGHTSM14}:
different moments of $h$ will scale with different exponents.)
Here $\zeta$ should be uniquely determined by which universality class the system belongs to. However, in practice, the observed $\zeta$ varies (See table~\ref{table:roughness}) even for the same type of system, such as paper wetting. Measuring a single exponent for these systems may prove inadequate due to the presence of
{\em crossover behavior} between universality classes. This is a common
source of confusion and controversy. If the crossover is gradual, an 
experiment or simulation may measure an effective exponent $\zeta_{eff}$ 
intermediate between existing theories, and appear to demand a new 
theoretical explanation ({\em i.e.}, universality class).

\begin{table*}[htp]
\begin{center}
\begin{tabular}{c c c}
Experiment & $\zeta$ & Reference \cr
\hline
\hline
Fluid flow & 0.73 & \cite{RubioPRL1989}\cr
& 0.81 & \cite{HorvathJPhysA1991} \cr
& 0.65-0.91 &  \cite{hePRL1992}\cr
\hline
Paper wetting & 0.63& \cite{Buldyrev1992PRA}\cr
&0.62-0.78& \cite{Family1992}\cr
\hline
Bacteria growth & 0.78 & \cite{vicsek1990physicaA}\cr
Burning fronts & 0.71 & \cite{zhang1992physicaA}\cr
\hline
\hline
\end{tabular}
\end{center}
\caption{\textbf{Roughness Exponents in Experiments.} Table reproduced from~\cite{barabasi95fractal}. Notice that there is a wide range of $\zeta$ reported, even for the same experimental system.}
\label{table:roughness}
\end{table*}

In section~\ref{sec:Fracture} we analyze a straightforward experimental example
of a crossover between two forms of roughness in two-dimensional fracture.
There we introduce the universal crossover scaling functions, 
and provide a brief renormalization-group rationale.

In the remainder of the paper, we examine a more complex theoretical model. 
Crossovers, long studied in ordinary critical phenomena,
have now been studied for several interface models~\cite{Buldyrev1992PRA}, however theoretical studies have proven challenging in different ways~\cite{anisotropicFRG}. For thin film magnets, the experiments~\cite{KimPRL03,ryu07nature,Sethna07Crossover} observe a crossover between short-range and mean-field universality classes as long-range dipolar fields are introduced, which can be done by changing the thickness of the film. 
However, for models of that type, simulations are challenging, both because of the long-range fields and the striking zig-zag morphologies that emerge and compete with the avalanche behavior. Crossovers involving the transition 
between depinning and sliding dynamics incorporating periodically correlated
disorder~\cite{BustingorryKG10} have also been studied. It is not typical,
however, to study and report the universal scaling functions for these
crossovers -- a challenge we now shall address.

We shall analyze a numerically tractable, but analytically tricky crossover~\cite{anisotropicFRG}: the transition between the linear, super-rough, quenched Edwards-Wilkinson model (qEW) and the nonlinear quenched KPZ model (qKPZ)~\cite{kardar97interfaces,Buldyrev1992PRA}. In both experiment and
theory, we focus on the crossover behavior of the height-height correlation function. 

\section{Crossover in fracture surface correlations}
\label{sec:Fracture}

\begin{figure}[htbp]
   \begin{center}
   \includegraphics[scale=0.6]{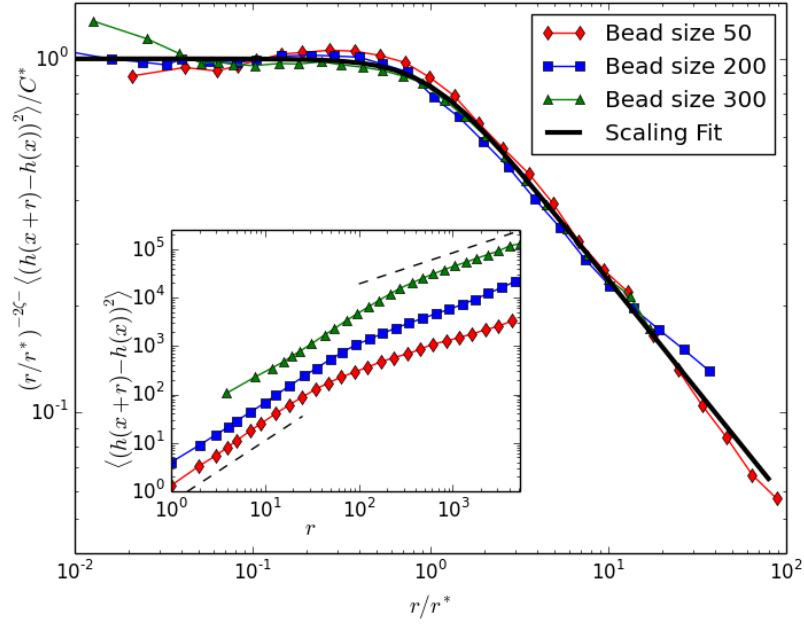}%
   \end{center}
   \caption{(Color online.) \textbf{Crossover scaling in fracture
   roughness~\cite{SantucciGHTSM14}.} The inset shows experimental data
   for the height-height correlation function
   $C(r) = \langle (h(x+r) - h(x))^2\rangle$ of a 2D fracture front, generated
   by pulling apart two pieces of PMMA that have been sand-blasted and 
   sintered together~\cite{SantucciGHTSM14}. The three curves differ in the
   size of the sand-grain beads; the relation between the bead size and the
   toughness fluctuations in the PMMA were not measured. The dashed lines
   show two different power-law critical regimes, with
   $C(r)\sim r^{2 \zetaMinus}$ and $C(r)\sim r^{2 \zetaPlus}$, governing the
   short- and long-distance scaling behavior: the crossover between these
   regimes is evident. Here our fit gives
   $\zetaMinus = 0.63$ and $\zetaPlus=0.32$, within the experimentalists
   suggested range $\zetaMinus = 0.6\pm0.05$ and $\zetaPlus=0.35\pm0.05$.
   The main figure shows a scaling plot of $r^{-2 \zetaMinus} C(r)$
   versus $r$, with the curves shifted vertically and horizontally to best
   collapse. The thick black curve is a one-parameter fit of the universal
   scaling function to the functional form in eqn~(\ref{eq:FractureCScaling}).
   }
\label{fig:fracture}
\end{figure}

Just as the critical exponent $\zeta$ is universal (independent of microscopic
details, within a class of physical system), so too is the crossover behavior
between universality classes.
As a simple example, Santucci et al.~\cite{SantucciGHTSM14} have measured
a relatively sharp crossover between two regimes for two-dimensional fracture
(inset in Fig.~\ref{fig:fracture}).  
Because the fracture is done slowly, we can view the crack front as self-organizing
to the depinning transition for the crack front.
Well below a critical distance $r^*$, 
they observe a power law $C(r) \sim r^{2\zetaMinus}$ with an exponent that was interpreted
as originating from coalescing cracks~\cite{SchmittbuhlCoalesce} or with Larkin scaling \cite{laurson2010}.
Well above $r^*$ they observe a different power law $C(r) \sim r^{2\zetaPlus}$ consistent with 
the depinning transition of a line~\cite{SchmittbuhlFlucLine,RussoK02,laurson2010}. 
The crossover between these two universal power-law regimes should be
described by a {\em universal crossover function}~\cite{Nelson75},
$\calCfrac$:
\begin{equation}
\label{eq:CFracScaling}
C(r) \approx C^* r^{-2 \zetaMinus} \calCfrac(r/r^*)
\end{equation}
independent of microscopic details. At small arguments $\calCfrac(X)$ must
go to a constant, and at large arguments 
$\calCfrac(X) \sim X^{2 (\zetaMinus-\zetaPlus)}$, so as to interpolate
between the two power laws.
When analyzing different systems
governed by the same universal crossover, one may plot all the crossovers
in a scaling plot, dividing the distances $r$ on the ordinate by a
system-dependent factor $r^*$
for each curve, and dividing the magnitudes of the correlations on the abscissa
by a system dependent constant $C^*$ (see Fig.~\ref{fig:fracture}). The
resulting data curves then should align, giving the universal function
$\calCfrac(r/r^*)$.

To continue with this simple test case, we may fit the universal scaling
function to an approximate functional form. (Indeed, we find it convenient
to do a joint fit of the functional form, the exponents, and the constants
$r^*$ and $C^*$.) To the extent that a guessed functional form reproduces
the universal one, it is equivalent: advanced field-theoretic methods for
calculating exact scaling functions aren't needed to analyze future experiments.
However, judicious choices of functional forms with the correct limiting
behavior can greatly facilitate this process. The interpolation
$1/(1+X^{-2(\zetaMinus-\zetaPlus)})$ has the correct limits, but its
rather gradual
crossover does not explain the data. We may heuristically add a parameter
$n$ which at large values produces an abrupt crossover:
\begin{equation}
\label{eq:FractureCScaling}
\calCfrac(X) = (1 + (X^{2 (\zetaMinus-\zetaPlus)})^{-n})^{-1/n}.
\end{equation}
This yields an excellent fit to the data with $n\approx 4$
(see Fig.~\ref{fig:fracture}). 

Why is the scaling form of eqn~(\ref{eq:CFracScaling}) expected? Briefly,
the renormalization group studies the behavior of systems under
coarse-graining: describing the properties of a system at length scales
changed by a factor $b$. One gets universal power laws when the system
becomes invariant under repeated coarse-grainings: if
$C(r) \to b^{2\zeta} C(r/b)$,
under coarse-graining by a factor $b$, then by coarse-graining $n$ times
such that $r = b^n$ one has $C(r) \propto b^{2\zeta n} = r^{2\zeta}$.
In the case of a crossover, a fixed point is unstable to some direction 
$\lambda$ in system space. Then a small initial $\lambda$ grows under 
rescaling by some factor $b^{1/\phi}$, so 
$C(r, \lambda) \to b^{2\zetaMinus} C(r/b, \lambda b^{1/\phi})$. Now
rescaling until $b^n = r$, we have 
\begin{equation}
C(r,\lambda) \to b^{2 n \zetaMinus} C(r/b^n, \lambda b^{n/\phi})
		= r^{2\zetaMinus} C(1, \lambda r^{1/\phi})
		= r^{2\zetaMinus} \calCfrac(\lambda^\phi r)
\end{equation}
where we choose $\calCfrac(X) = C(1,X^{1/\phi})$. If the unstable
direction flows to a new fixed point with a different $\zetaPlus$, 
that behavior will be reflected in the large-X dependence 
$C(X)\sim X^{2(\zetaMinus-\zetaPlus)}$~\cite[Section 4.2]{Cardy96}. 
Note that different
physical systems will have different overall scales of height fluctuations,
so we must have an overall scale $C^*$ for each experiment. (If the 
experiments fall into a parameterized family, $C^*$ will depend 
smoothly on the parameters, giving {\em analytic corrections to scaling}
as discussed in Section~\ref{sec:Results}.)
Note, though,
that the rescaling factor $r^*$ for lengths, while it will still vary
from one system to another, now depends on $\lambda$ with a power-law
singularity, as $r^* = 1/\lambda^\phi$; within the renormalization group, 
$\lambda$ measures
how far along the unstable direction the original system was poised.
In particular, $r^*$ becomes large as $\lambda\to 0$, as
in that limit the unstable fixed point remains in control.

The three experiments depicted in Fig.~\ref{fig:fracture} started with
different bead sizes. If all other features of the experiment are held fixed,
one may assume that the control parameter $\lambda$ depends in some
smooth way on the bead size. Had we several values of bead size,
we could then extract values for the universal crossover exponent $\phi$.

In the following sections, we shall perform a far more sophisticated
version of this type of analysis. By exhaustively varying system size
and nonlinearity 
in an interface growth model, we shall not only generate universal
{\em two-variable} functional forms for the 
correlation crossover scaling function, but will be able to make predictions
about both the dependence of the crossover length scale (corresponding to
$r^*$) and the dependence of the correlation amplitude (corresponding to 
$C^*$) on the control parameters. A rich, nuanced understanding of the
model behavior thus emerges.

\section{Line depinning model}
\label{sec:Model}

The equations of an interface in a disordered environment may be written generally as follows.  Let the one-dimensional interface, $h(x,t)$ be driven by a force $H(t)$ through a disordered environment with a local quenched random force $\eta(h(x),x)$:
\begin{equation}
\label{eq:FrontProp}
\frac{\partial h}{\partial t}  = \gamma \nabla^2 h + \lambda (\nabla h)^2
			+ \eta(h(x),x) - k \langle h \rangle_x + H(t).
\end{equation}
Here $\gamma$ is a surface tension, and $\lambda$ is the coefficient of the
KPZ term. The KPZ term controls lateral spreading of the interface, breaks the 
statistical tilt symmetry, and changes the universality
class~\cite{barabasi95fractal}.
$H(t)$ is a slowly-increasing external driving force.
Our simulations are done with a lattice
automaton; the lattice naively might be thought to break this statistical
tilt symmetry, but simulations have long shown that the model faithfully 
describes both universality classes~\cite{amaral95}.

The term $-k \langle h \rangle_x$ is borrowed from simulations of magnets,
where it represents the demagnetization force~\cite{DurinZ06},
approximating the effects of the long-range dipolar field cost of a net 
advance in the front. This restoring force `self-organizes' the depinning transition to the fixed point, allowing simulations to access many metastable states,
without having to enforce an actual quasi-static field. 
It is known~\cite{narayan96} that this restoring force does not produce loop corrections to the renormalization group equations and therefore
does not change the universality class of the problem.
We have confirmed numerically that its effects are small for our crossover
and appear irrelevant. As the restoring force makes the simulation 
vastly more efficient, we include this restoring force, but we do not
include $k$ in our scaling analysis.



\section{Analysis of crossover scaling}
\label{sec:Results}

\begin{figure}[htbp]
   \begin{center}
   \includegraphics[scale=0.6]{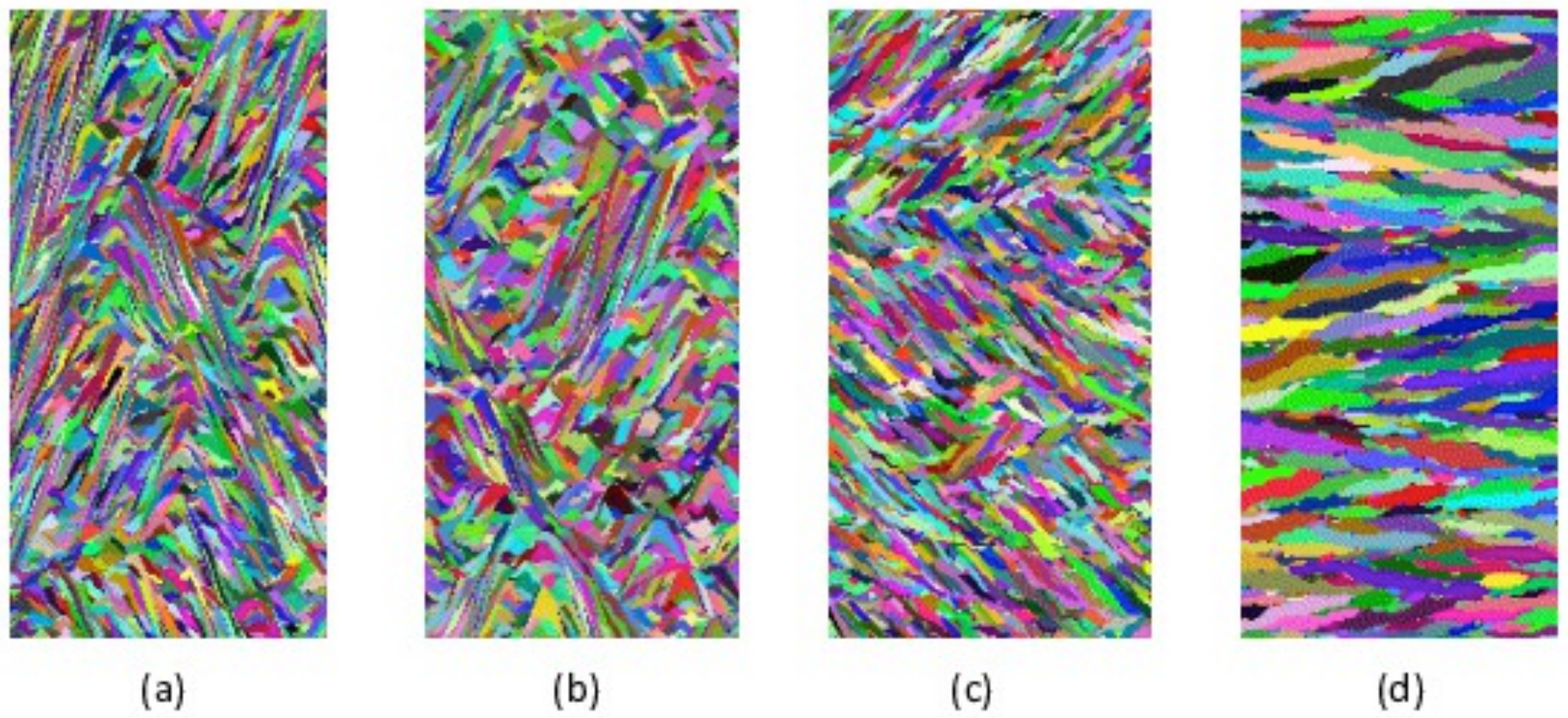}%
   \end{center}
   \caption{(Color online.)
   \textbf{Crossover of qKPZ to qEW Model.} Fronts generated from
   128$\times$256
   simulations with the nonlinear KPZ term coefficients set to
   (a)~$\lambda=0$, (b)~$\lambda=0.001$, (c)~$\lambda=0.1$, (d)~$\lambda=5$.  The
   random colors/shades represent the area between each pinned front. 
   One can see that the morphology of the interfaces change
   dramatically as $\lambda$ increases.}
\label{fig:crossover}
\end{figure}

Using the automaton simulation employed in~\cite{ChenPSZD11}, we tune $\lambda/\gamma$ from 0 to 5, and observe how the resulting behavior changes. Figure~\ref{fig:crossover} shows how the front morphology qualitatively changes while we increase the nonlinear parameter $\lambda$. Notice that with increasing $\lambda$ the fronts between events are flatter than at small $\lambda$. 

\begin{figure}[htbp]
   \begin{center}
   \includegraphics[scale=0.5]{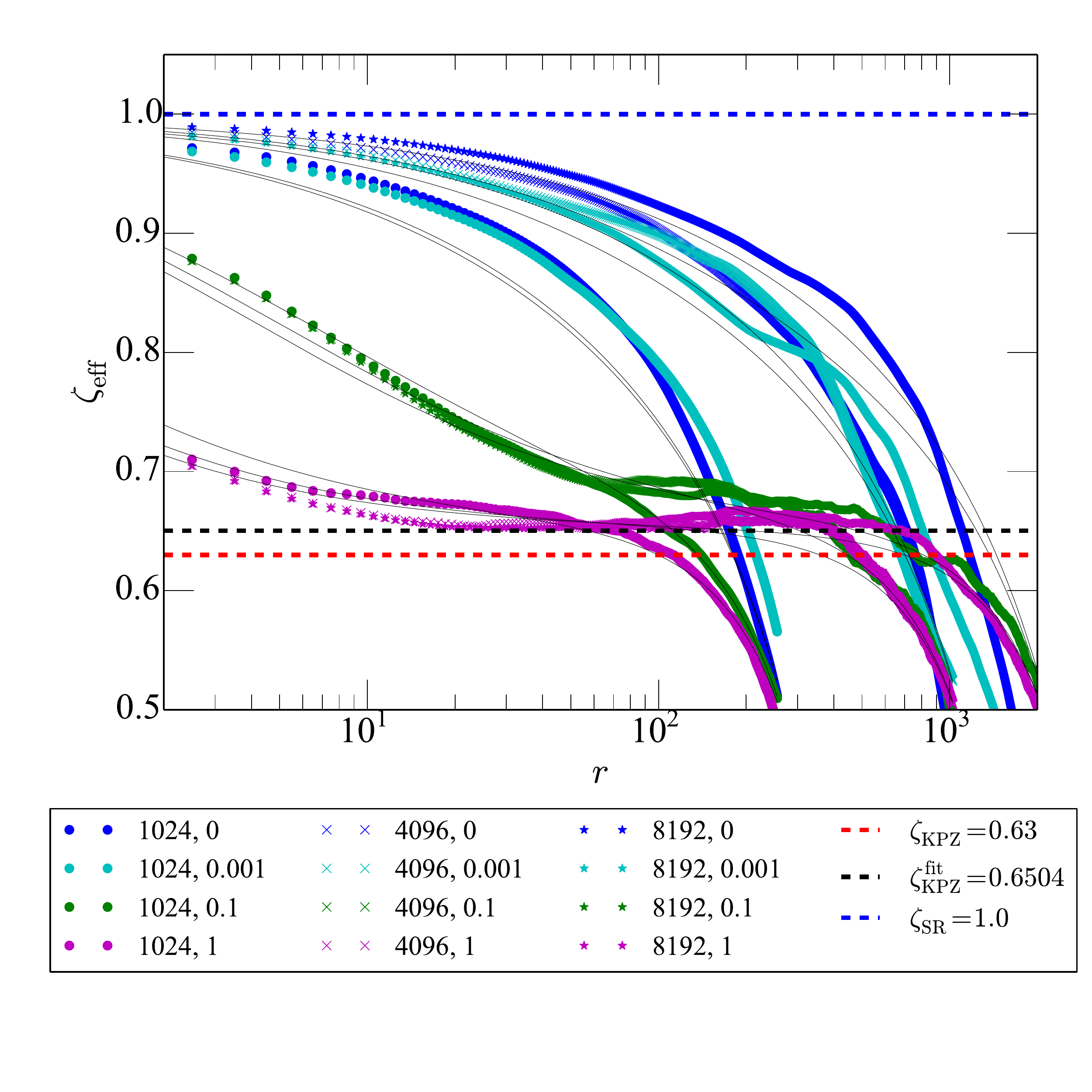}%
   \end{center}
   \vskip -0.9truein
   \caption{(Color online.)
   \textbf{Local Log Slope.} The measured local-log slope,
$d\log{C}/d\log r$ of the height-height correlation function for varying
$\lambda$ and $k$.
The lower dashed red line is $\zetaKPZ=0.63$ as expected for the KPZ
universality class. The upper dashed blue line is $\zeta_{SR} = 1.0$, the
largest growth allowed for super-rough interfaces. The thin lines
show the predictions of our fits (Section~\ref{sec:Results}),
and the dashed black line is the fit value of $\zetaKPZ$.
The sharp cutoff in the curves at large $r$ is due to the periodic boundary
condition, that forces $\zeta_{eff} = 0$ at $r=L/2$; the larger $L$ curves
have later cutoffs. Note as expected that for small $\lambda$ the curves
are described by the EW behavior $\zeta_{SR}=1$ at small $r$, while for
large $\lambda$ they are well described by $\zetaKPZ$. For
intermediate values of $\lambda \sim 0.1$, we observe a clear transition
from EW behavior at small $r$ to $KPZ$ behavior at intermediate $r$, before
being cut off by the finite size effects.}
\label{fig:local_log_new}
\end{figure}

According to Equation~\ref{eq:heightzeta}, naively one would assume we could recover the exponent $\zeta$ by defining an effective exponent $\zeta_{eff}$ to be half the local-log slope of the height-height correlation functions
(Figure~\ref{fig:local_log_new}). From other numerical studies, for qEW,
we expect $\zetaEW \simeq 1.25$ (Cellular automata~\cite{leschhorn93,JostautomatonqEW97} models show $\zetaEW=1.25\pm0.01$; continuous string models~\cite{FRGDepinningRosso}
found $\zetaEW \approx 1.26$.). For qKPZ, we expect $\zetaKPZ=0.63$~\cite{leschhorn96,LeeK05}. However, there are two things about Figure~\ref{fig:local_log_new} worth noting: (1)~the slope-measure of $\zeta$ drifts between $0.63$ and $1.0$ as we change $\lambda$, (2)~the measured value is never greater than one as is naively expected for the linear qEW model. The dropoff at $r\sim L/4$
is due to the periodic boundary conditions.

The second issue has a known resolution: for $\zeta > 1$, when the interface is `superrough', the height cannot grow faster than linearly with distance, so the height-height correlation function cannot directly exhibit a power law larger than one~\cite{PRLZetaComment}.
This so-called anomalous scaling~\cite{LopezRC97,LopezS98,Ramasco00} implies
that the exponent $\zeta$ is reflected not in the distance-dependence of the
correlation function, but rather in its {\em system-size dependence}.
We thus consider the finite-size scaling form
\begin{equation}
\label{eq:Cew}
C_{EW}(r|L) \sim L^{2 \zetaEW} (r/L)^2 {\cal C}_{EW}(r/L);
\end{equation}
the roughness exponent $\zetaEW$ may be estimated by the system-size
dependence of the magnitude of $C_{EW}$. Note that the periodic boundary
conditions implies that $C_{EW}(r|L) = C_{EW}(-r|L) = C_{EW}(L-r|L)$; near
$r = L/2$ the correlation function reaches a peak (and $\zeta_\mathrm{eff}$
vanishes, as in Fig.~\ref{fig:local_log_new}). Thus 
$X^2 {\cal C}_{EW}(X) = (1-X)^2 {\cal C}_{EW}(1-X)$. To control the sharpness
of the peak in the correlation function at $X=1/2$, in analogy to the
crossover sharpness parameter $n$ of eqn~\ref{eq:FractureCScaling}, we
introduce $n_{EW}$ giving a transition between the two power-laws:
\begin{equation}
\label{eq:CewForm}
X^2 {\cal C}_{EW}(X) = ( (X^2)^{-n_{EW}} + ((1-X)^2)^{-n_{EW}} )^{-1/n_{EW}}.
\end{equation}
For qKPZ (Fig.~\ref{fig:crossover}c), the correlation function in a system size $L$ takes the finite-size scaling form 
\begin{equation}
\label{eq:Ckpz}
C_{KPZ}(r|L) = A r^{2\zetaKPZ} {\cal C}_{KPZ}(r/L),
\end{equation}
where we introduce $n_{KPZ}$ to form a periodic functional form
\begin{equation}
\label{eq:CkpzForm}
X^{2 \zetaKPZ} {\cal C}_{KPZ}(X) 
   = ( (X^{2 \zetaKPZ})^{-n_{KPZ}} 
	+ ((1-X)^{2 \zetaKPZ})^{-n_{KPZ}} )^{-1/n_{KPZ}}.
\end{equation}

\begin{figure}[htbp]
   \begin{center}
   \includegraphics[scale=0.6]{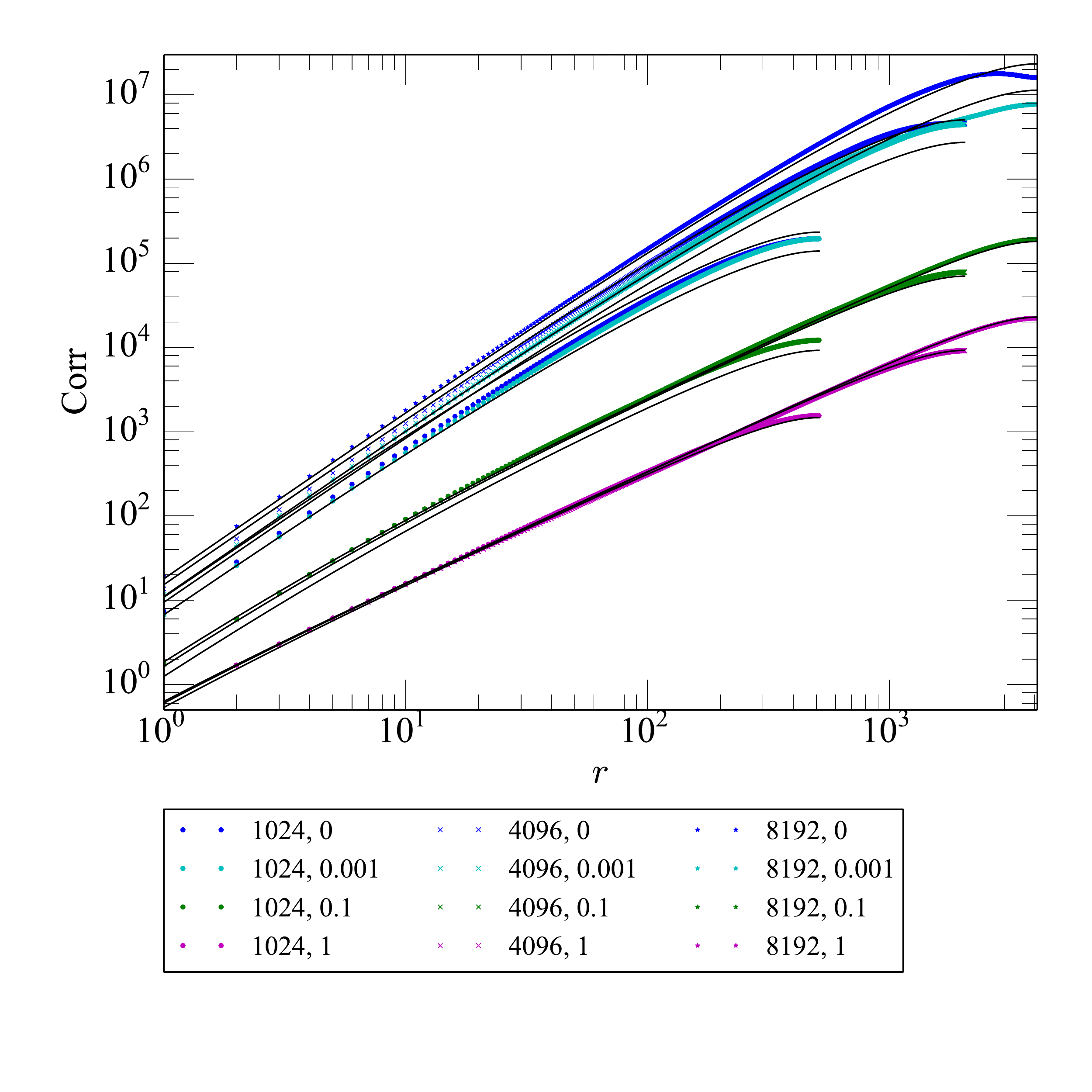}%
   \end{center}
   \vskip -0.5truein
   \caption{(Color online.)
\textbf{Height-height Correlation Function.} The numerics generated with
an automata code (symbols) are well described by eqn~\ref{eq:corrcrossover}
(black curves) with fit parameters
$\phi= 1.0 \pm 0.4$,
$\zetaKPZ= 0.65 \pm 0.04$,
$\zetaEW= 1.1 \pm 0.15$,
$n_{Cross}= 1.0 \pm 0.7$,
$B= 2.5 \pm 6.0$.
$n_{EW}= 0.27 \pm 0.03$,
$n_{KPZ}= 1.1 \pm 0.6$,
$A_0= 2.7 \pm 7$,
$A_1= 770 \pm 900$, and
$A_\infty = 0.3 \pm 1$.
(A fit constrained to the expected values of $\zetaEW$ and $\zetaKPZ$ give
slightly worse, but acceptable fits, with the other parameter estimates
within the quoted ranges.)
The legends denote $L$ and $\lambda$ for each
simulated correlation function; all runs had $k=0.01$.
The errors quoted are a rough measure of the systematic
error~\cite{FrederiksenJBS04}, as described in the text, and
are representative of the differences we find using different weightings
and functional forms. (They are much larger than the statistical errors.)
Note that the amplitude dependence is captured by the scaling form. 
Three of the twelve parameters ($\zetaEW$, $\zetaKPZ$, 
and $\phi$) are universal critical exponents,
three ($A_0$, $A_1$, and $A_\infty$) describe the non-universal dependence
of an overall height scale on parameters, two describe finite-size effects,
and only two ($n_{Cross}$ and $B$) are needed to describe the universal
crossover function to the accuracy shown.}
\label{fig:corrFit}
\end{figure}

\begin{figure}[htbp]
   \begin{center}
   (a) \includegraphics[scale=0.28]{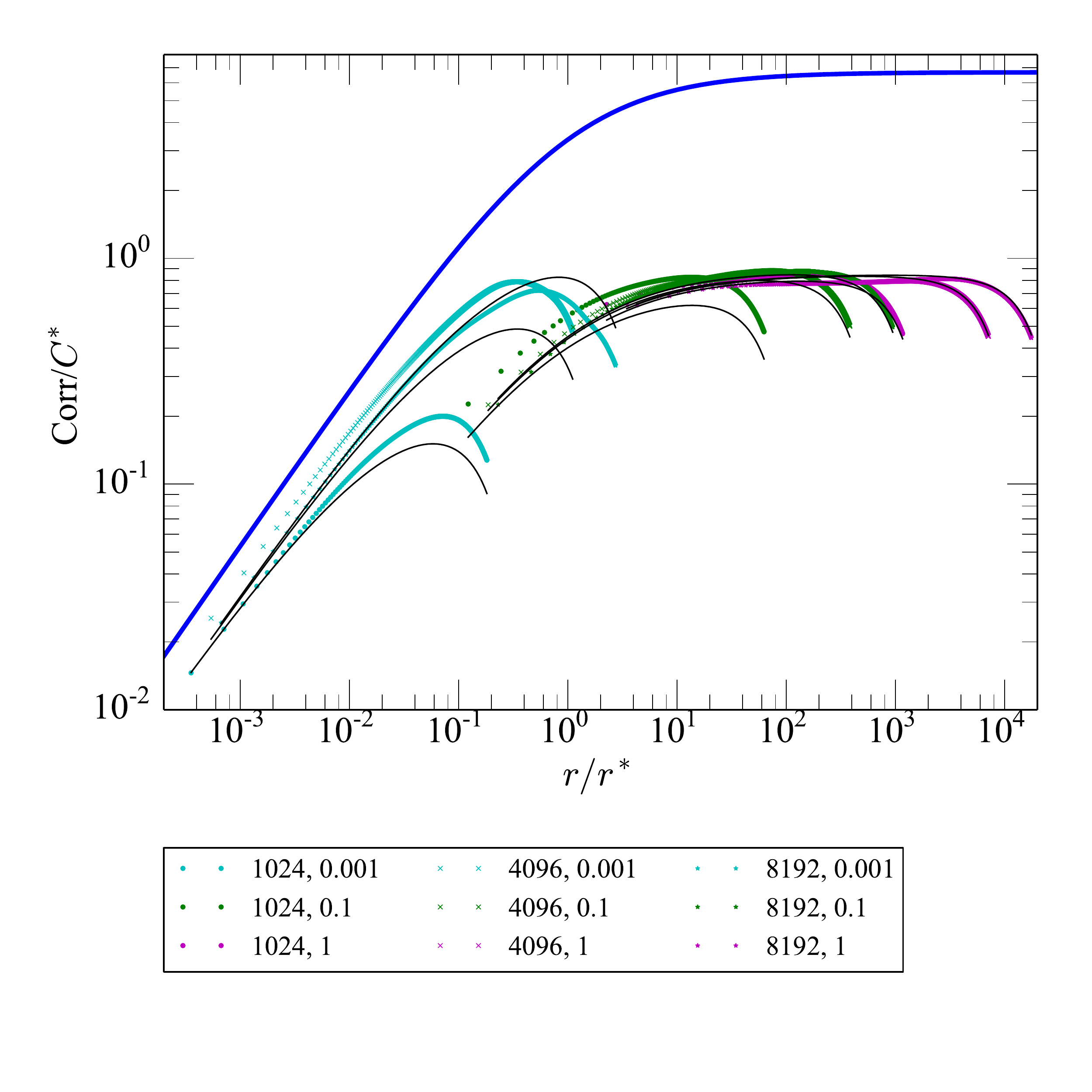}%
   (b) \includegraphics[scale=0.28]{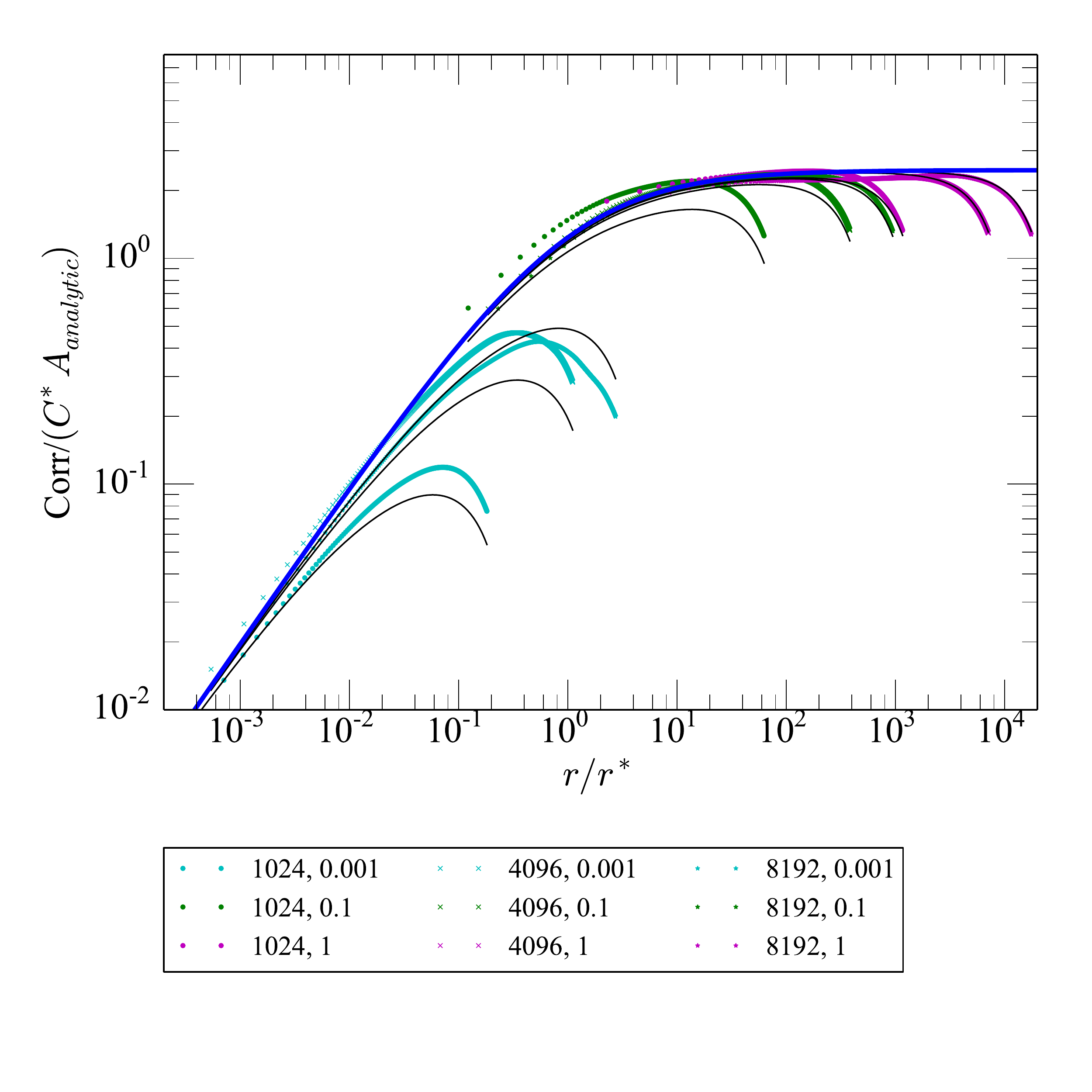}%
   \end{center}
   \vskip -0.3truein
   \caption{(Color online.)
\textbf{Scaling collapse of the height-height correlation function.} 
We collapse the correlation-function data of Fig.~\ref{fig:corrFit} to 
illustrate the crossover between EW and KPZ-dominated lengths
($r<r^*$ and $r>r^*$ respectively, see also~\cite{ryu07nature,Sethna07Crossover}). Here 
$r^* = L [B (L \lambda^\phi)^{2 (\zetaEW - \zetaKPZ)} ]^{-1/(2-2\zetaKPZ)}$ is the distance where the EW and KPZ components of the correlation
function are equal in magnitude, and $C^* = \lambda^{-2 \phi (\zetaEW-\zetaKPZ)} r^{2 \zetaKPZ}$ factors out the dependence expected in the KPZ regime
(hence yielding flat behavior for $r\gg r^*$).
In (b), we also factor out the effects of analytic corrections
to scaling; here all of the curves lie on a universal curve apart from the
effects of the finite system sizes (causing each curve to drop on the right).
In (a), we see that the analytic correction to scaling has a significant
impact on the scaling collapse: ignoring it in the analysis would produce
significant errors in critical exponents and scaling functions. The thick
blue curve shows the scaling function prediction for the crossover
(which, if $x = r/r^*$, can be shown to be $C/C^* = B x^2 (
x^{2 n_{Cross}} + x^{2 n_{Cross}\zetaKPZ})^{-1/n_{Cross}}$).
}
\label{fig:corrCollapse}
\end{figure}

The drift in the exponent $\zeta$, however, demands a study of the 
scaling near the unstable qEW fixed point, and the functional form of the
resulting crossover scaling function.
The role of $\lambda$ in generating the crossover from qEW to qKPZ has only been studied qualitatively~\cite{tang92, leschhorn94,leschhorn96,rosso01}, with no full description of the crossover scaling~\cite{anisotropicFRG}. 
The crossover describes the RG flow from the qEW fixed point to the qKPZ as the relevant parameter $\lambda$ is added. The scaling form for the height-height correlation function is thus that of a relevant variable $\lambda$ added to the qEW scaling:
\begin{equation}
\label{eq:Ccross}
C(r|L,\lambda) = L^{2 \zetaEW} {\cal C}(r/L, \lambda^\phi r).
\end{equation}
For $\lambda \gg 0$, $C(r |L, \lambda) \rightarrow C_{KPZ}(r|L)$, therefore,
\begin{align}
{\cal C}(r/L, \lambda^\phi r) &\to A(\lambda) r^{2 \zetaKPZ} {\cal C}_{KPZ}(r/L) / L^{2 \zetaEW} \cr
& = A(\lambda) (r/L)^{2 \zetaKPZ} L^{-2 (\zetaEW-\zetaKPZ)} {\cal C}_{KPZ}(r/L) \cr
& = A_{analytic}(\lambda) (r/L)^{2 \zetaKPZ} (\lambda^\phi L)^{-2 (\zetaEW-\zetaKPZ)}  {\cal C}_{KPZ}(r/L).
\label{eq:crossCurlyC}
\end{align}
Here $A(\lambda)$ is in general a non-universal prefactor for the $KPZ$ 
correlation function. As with $C^*$ in Section~\ref{sec:Fracture}, $A$ is
expected to vary%
  \footnote{There are other analytic corrections to scaling that will
  in general become important at long distances from the critical point. 
  For example~\cite[Section 3.3]{Cardy96},
  $\lambda$ is presumably not the natural measure of 
  the unstable direction in parameter space; in general 
  that would be an analytic function $u_\lambda \propto \lambda + O(\lambda^2)$
  of $\lambda$ and other parameters.}
with the parameters of the problem. Here it has a typical
smooth variation $A_{analytic}(\lambda)$, times a singular piece 
$A_{singular}(\lambda)$ due to the EW fixed point:
$A(\lambda) = A_{analytic}(\lambda) A_{singular}(\lambda)$. We
derive the power-law divergence of the amplitude 
\begin{equation}
\label{eq:Asingular}
A_{singular}(\lambda) = \lambda^{-2(\zetaEW-\zetaKPZ) \phi}
\end{equation}
by noting that ${\cal C}(r/L, \lambda^\phi r)$ must be a scaling function 
with only invariant combinations of $r$, $L$, $\lambda$. 

We must also have ${\cal C}(r/L, 0) \sim (r/L)^2 C_{EW}(r|L)$.
Using these limits, Equation~\ref{eq:crossCurlyC}, and the fact that
$A(\lambda)$ gets large as $\lambda \to 0$ and small as $\lambda \to \infty$,
we can construct a function that crosses over between these two limits:
\begin{align}
\label{eq:corrcrossover}
C(r|L, \lambda)
=& A_{analytic} * 
  \left((A_{singular}(\lambda) B C_{KPZ}(r/L))^{-n_{Cross}}+C_{EW}(r/L)^{-n_{Cross}}\right)^{-1/n_{Cross}}\cr
=& L^{2 \zetaKPZ} {\cal C}(X,Y),
\end{align} 
where $X=r/L$ and $Y=\lambda^\phi r$, and $C_{KPZ}$ and $C_{EW}$ are defined
in eqns.~(\ref{eq:Cew}-\ref{eq:CkpzForm}). We
expand the analytic function $A_{analytic} = (A_0 - A_\infty)/(1 + A_1 \lambda) + A_\infty$ in a form analytic at zero and saturating at large $\lambda$ at
$A_\infty$,%
   \footnote{Indeed, simulations at $\lambda = 2$ and $\lambda = 5$ 
   indicate a saturation of the amplitude of the KPZ power-law scaling.
   This suggests that $A_{analytic}(\lambda) \propto 1/A_{singular}(\lambda)$
   at larger $\lambda$. In the range we consider, an asymptotically
   flat asymptote suffices to capture this variation; independent fits
   of the amplitude at each $\lambda$ including the larger systems 
   give comparable fits and other parameters within our ranges.}
and we include a relative scale
factor $B$. Finally, we vary the sharpness of the crossover with $n_{Cross}$,
just
as we did in the experimental study of fracture (eqn~\ref{eq:FractureCScaling}).

\begin{figure}[htbp]
   \begin{center}
   \includegraphics[scale=0.6]{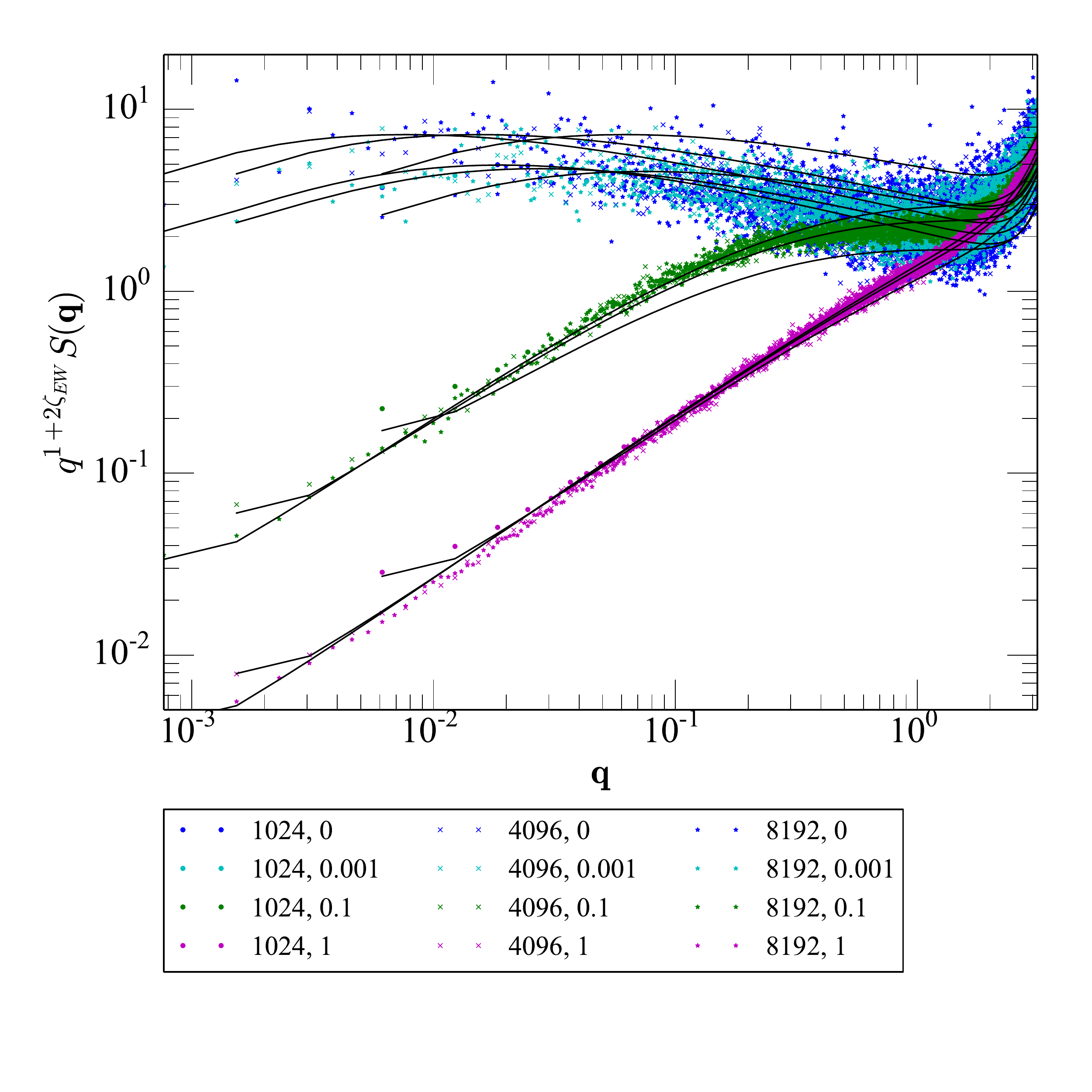}%
   \end{center}
   \vskip -0.5truein
   \caption{(Color online.)
\textbf{Rescaled spectral function.} This is the spectral function
$S(q| L, \lambda)$ for the height fluctuations. Note that rescaling
by the naive RG power $q^{1+2 \zetaEW}$ here does collapse the data
for vanishing $\lambda$ (up to a significant noise): the anomalous scaling
for the super-rough interface in real space is not manifested in Fourier
space. The theory curves are given by the Fourier transform of the same
best fit shown in previous figures.}
\label{fig:Sofq}
\end{figure}

The theoretical curves in
were fit to the data in Figs.~(\ref{fig:local_log_new}) and~(\ref{fig:corrFit})
and~(\ref{fig:Sofq}),
deleting the noisy half near $r=L/2$ in the first, and
using weights $\sigma^2 \sim \sqrt{r/L}$ designed to equalize the emphasis
on each decade. The errors quoted are a rough estimate of the systematic
error~\cite{FrederiksenJBS04} given by quadratically exploring fits with 
roughly twice the $\xi^2$ of the best fit.

Note that this gives us a universal function of three variables
($r$, $L$, and $\lambda$). Note that it predicts a singularity at small 
$\lambda$ in the form of a divergent
amplitude $A_{sing} \sim  \lambda^{-2(\zetaEW-\zetaKPZ) \phi}$
in the qKPZ correlation function (eqn~\ref{eq:Ckpz}) as $\lambda\to0$.
This universal
singularity in the amplitude (corresponding to a prediction of $C^*$ in
section~\ref{sec:Fracture}) explains the amplitude dependence seen in
Fig.~\ref{fig:corrFit}. There is an analogous universal amplitude
dependence seen for the Heisenberg$\to$Ising crossover at small Ising
anisotropy~\cite[Section 4.1]{Cardy96}.

We can use the scaling form of the correlation function $C(r|L, \lambda)$ to 
derive other, less complex crossover scaling functions traditionally studied
in interface depinning problems~\cite{BustingorryKG10,FRGDepinningRosso}.
The spectral scaling function is equal to the
Fourier transform of our correlation function (with subtleties at $q=0$):
\begin{equation}
\label{eq:Sofq}
S(q|L,\lambda) = |{\tilde h}(q)|^2 = \int_r C(r|L,\lambda) \exp(i q r) dr
	\sim q^{-1-2\zetaEW} {\cal S}(\lambda^{-\phi} q, q L).
\end{equation}
Here the universal spectral crossover scaling function
${\cal S}(\tilde X, \tilde Y)$ can be written in terms of our universal 
correlation crossover scaling function (${\cal C}(X,Y)$ from
eqn~\ref{eq:Ccross}) as
\begin{equation}
\label{eq:SpectralUniversal}
{\cal S}(\tilde X, \tilde Y) 
	= \int dz \exp(i z) {\tilde Y}^{2 \zetaEW}
			{\cal C}(z/{\tilde Y}, z/{\tilde X})
\end{equation}
depicted in Fig.~(\ref{fig:Sofq}). Here it is known that ${\cal S}$ does
not have anomalous scaling; the power laws $\zetaEW$ and $\zetaKPZ$
can be read off from the slopes at large and small $\tilde Y$. 
More ordinary, single-variable scaling functions can be derived from
our two-variable scaling functions, such as that governing the
average width of an interface (given by the zero Fourier component of
$S(q)$ or an integral of $C(r)$; such scaling functions 
(as for the fracture scaling function of eqn~\ref{eq:CFracScaling})
allow traditional scaling collapses (as in Fig.~\ref{fig:fracture}).
However, one should note that our analyzed simulations extend to 
$\lambda\sim 1$, where analytic corrections to scaling, as vividly illustrated
in Fig.~\ref{fig:corrCollapse}(a), would likely invisibly distort the
resulting scaling collapses. It is an advantage of multivariable scaling 
fits that they both allow the incorporation of such analytic corrections
(extending the range of applicability), and force their incorporation
(exposing weaknesses of the naive theory). 


\section{Conclusions}
\label{sec:Conclusions}

In this paper, we have analyzed the scaling properties for an experimental
2d fracture front and a model of an interface moving in random media,
focusing on the crossover scaling of the roughness. The experimental
system is successfully modeled using a one-variable universal scaling function
with one free parameter, controlling the sharpness of the transition.
The theoretical model, the crossover from the qEW to the
qKPZ universality class with the addition of a non-linear term, allows us
to estimate the complete universal scaling function for the
height-height correlation function including both finite-size effects and the
non-linear effects of the tuning parameter $\lambda$, while satisfying known
limits given by the renormalization group. 

We emphasize here the importance of our sophisticated use of the scaling
forms and corrections predicted from the renormalization group. 
Fig.~\ref{fig:fracture} illustrates that not only the power-laws, but the
entire functional form of the crossover, is a universal property that
should be reported. Equation~\ref{eq:FractureCScaling} is an effective
one-parameter way of embodying the sharpness of the crossover, which we 
use also in the theoretical analysis of Section~\ref{sec:Results} for
both the crossover and the effects of periodic boundary conditions.
Figures~\ref{fig:local_log_new}, \ref{fig:corrFit}, and~\ref{fig:Sofq}
show how different experimental characterizations of 
the roughness of an interface can be simultaneously fit with a single
functional form. Figure~\ref{fig:corrCollapse}(a) vividly indicates the
importance of analytic corrections to scaling in extending the validity
of the theoretical analysis to smaller systems and farther outside the
critical region. Only systems with $\lambda < 10^{-3}$ will follow the
scaling behavior without incorporating the analytic corrections, while
the entire crossover is faithfully represented in
Fig.~\ref{fig:corrCollapse}(b) by including them in the fit.

By developing functional forms for the correlation
functions~\cite{ChenPSZD11}, we gain the
flexibility of incorporating analytic corrections, multiple scaling variables,
and a systematic error analysis while allowing the quantitative reporting
of the universal scaling functions. One should note that the parameter
estimation errors quoted here are large compared to more traditional 
scaling analyses. In part this is due to our estimation of the relevant
systematic errors~\cite{FrederiksenJBS04}; statistical errors would be
perhaps an order of magnitude smaller. In part, however, this is due to 
our incorporation of known but usually ignored confounding factors -- 
analytic corrections to scaling and crossover effects will invisibly
distort the results of more direct measurements, and the drift in exponents
quoted in the literature in critical phenomena often exceeds the error
estimates.

It is challenging but satisfying to develop these functional forms. 
Measuring and fitting them is far more physically intuitive and less
technically demanding than direct field-theoretic
calculations~\cite{FRGDepinningRosso} (although theoretical calculations often form
important inspiration for choosing functional forms~\cite{ChenPSZD11}). 
Successful functions are parsimonious in the number of adjustable parameters,
and developing them often forces one to
develop a far more complete understanding of the physics of the system
under consideration.

\section*{Acknowledgment}
This work is supported by NSF and CNR through Materials World Network: Cooperative Activity in Materials Research between US Investigators and their Counterparts Abroad in Italy (NSF DMR 1312160), and through NSF PHY11-25914. SZ acknowledges support from the Academy of Finland
FiDiPro program, project 13282993 and ERC advanced grant SIZEFFECTS..


\begin{thebibliography}{48}
\expandafter\ifx\csname natexlab\endcsname\relax\def\natexlab#1{#1}\fi
\expandafter\ifx\csname bibnamefont\endcsname\relax
  \def\bibnamefont#1{#1}\fi
\expandafter\ifx\csname bibfnamefont\endcsname\relax
  \def\bibfnamefont#1{#1}\fi
\expandafter\ifx\csname citenamefont\endcsname\relax
  \def\citenamefont#1{#1}\fi
\expandafter\ifx\csname url\endcsname\relax
  \def\url#1{\texttt{#1}}\fi
\expandafter\ifx\csname urlprefix\endcsname\relax\def\urlprefix{URL }\fi
\providecommand{\bibinfo}[2]{#2}
\providecommand{\eprint}[2][]{\url{#2}}

\bibitem[{\citenamefont{Barab{\'a}si and Stanley}(1995)}]{barabasi95fractal}
\bibinfo{author}{\bibfnamefont{A.~L.} \bibnamefont{Barab{\'a}si}}
  \bibnamefont{and} \bibinfo{author}{\bibfnamefont{H.~E.}
  \bibnamefont{Stanley}}, \emph{\bibinfo{title}{{Fractal concepts in surface
  growth}}} (\bibinfo{publisher}{Cambridge University Press},
  \bibinfo{year}{1995}), ISBN \bibinfo{isbn}{0521483182}.

\bibitem[{\citenamefont{{Kardar}}(1998)}]{kardar97interfaces}
\bibinfo{author}{\bibfnamefont{M.}~\bibnamefont{{Kardar}}},
  \bibinfo{journal}{Phys. Rep.} \textbf{\bibinfo{volume}{301}},
  \bibinfo{pages}{85} (\bibinfo{year}{1998}), \eprint{arXiv:cond-mat/9704172}.

\bibitem[{\citenamefont{Leschhorn et~al.}(1997)\citenamefont{Leschhorn,
  Nattermann, Stepanow, and Tang}}]{leschhorn97}
\bibinfo{author}{\bibfnamefont{H.}~\bibnamefont{Leschhorn}},
  \bibinfo{author}{\bibfnamefont{T.}~\bibnamefont{Nattermann}},
  \bibinfo{author}{\bibfnamefont{S.}~\bibnamefont{Stepanow}}, \bibnamefont{and}
  \bibinfo{author}{\bibfnamefont{L.~H.} \bibnamefont{Tang}},
  \bibinfo{journal}{Ann. Physik} \textbf{\bibinfo{volume}{6}},
  \bibinfo{pages}{1} (\bibinfo{year}{1997}).

\bibitem[{\citenamefont{Rosso and Krauth}(2001)}]{rosso01}
\bibinfo{author}{\bibfnamefont{A.}~\bibnamefont{Rosso}} \bibnamefont{and}
  \bibinfo{author}{\bibfnamefont{W.}~\bibnamefont{Krauth}},
  \bibinfo{journal}{Phys. Rev. Lett.} \textbf{\bibinfo{volume}{87}},
  \bibinfo{pages}{187002} (\bibinfo{year}{2001}).

\bibitem[{\citenamefont{Rosso and Krauth}(2002{\natexlab{a}})}]{rosso02}
\bibinfo{author}{\bibfnamefont{A.}~\bibnamefont{Rosso}} \bibnamefont{and}
  \bibinfo{author}{\bibfnamefont{W.}~\bibnamefont{Krauth}},
  \bibinfo{journal}{Phys. Rev. E} \textbf{\bibinfo{volume}{65}},
  \bibinfo{pages}{025101} (\bibinfo{year}{2002}{\natexlab{a}}).

\bibitem[{\citenamefont{Rosso et~al.}(2003)\citenamefont{Rosso, Hartmann, and
  Krauth}}]{rosso03}
\bibinfo{author}{\bibfnamefont{A.}~\bibnamefont{Rosso}},
  \bibinfo{author}{\bibfnamefont{A.~K.} \bibnamefont{Hartmann}},
  \bibnamefont{and} \bibinfo{author}{\bibfnamefont{W.}~\bibnamefont{Krauth}},
  \bibinfo{journal}{Phys. Rev. E} \textbf{\bibinfo{volume}{67}},
  \bibinfo{pages}{021602} (\bibinfo{year}{2003}).

\bibitem[{\citenamefont{Tang and Leschhorn}(1992)}]{tang92}
\bibinfo{author}{\bibfnamefont{L.-H.} \bibnamefont{Tang}} \bibnamefont{and}
  \bibinfo{author}{\bibfnamefont{H.}~\bibnamefont{Leschhorn}},
  \bibinfo{journal}{Phys. Rev. A} \textbf{\bibinfo{volume}{45}},
  \bibinfo{pages}{R8309} (\bibinfo{year}{1992}).

\bibitem[{\citenamefont{Leschhorn}(1993)}]{leschhorn93}
\bibinfo{author}{\bibfnamefont{H.}~\bibnamefont{Leschhorn}},
  \bibinfo{journal}{Physica A: Statistical and Theoretical Physics}
  \textbf{\bibinfo{volume}{195}}, \bibinfo{pages}{324 } (\bibinfo{year}{1993}),
  ISSN \bibinfo{issn}{0378-4371},
  \urlprefix\url{http://www.sciencedirect.com/science/article/B6TVG-46D5GTS-5C/2/4bf65953d8f7f821298068541bda4b36}.

\bibitem[{\citenamefont{Leschhorn}(1992)}]{leschhorn92}
\bibinfo{author}{\bibfnamefont{H.}~\bibnamefont{Leschhorn}},
  \bibinfo{journal}{J. Phys. A} \textbf{\bibinfo{volume}{25}},
  \bibinfo{pages}{L555} (\bibinfo{year}{1992}).

\bibitem[{\citenamefont{Leschhorn and Tang}(1994)}]{leschhorn94}
\bibinfo{author}{\bibfnamefont{H.}~\bibnamefont{Leschhorn}} \bibnamefont{and}
  \bibinfo{author}{\bibfnamefont{L.-H.} \bibnamefont{Tang}},
  \bibinfo{journal}{Phys. Rev. E} \textbf{\bibinfo{volume}{49}},
  \bibinfo{pages}{1238} (\bibinfo{year}{1994}).

\bibitem[{\citenamefont{Leschhorn}(1996)}]{leschhorn96}
\bibinfo{author}{\bibfnamefont{H.}~\bibnamefont{Leschhorn}},
  \bibinfo{journal}{Phys. Rev. E} \textbf{\bibinfo{volume}{54}},
  \bibinfo{pages}{1313} (\bibinfo{year}{1996}).

\bibitem[{\citenamefont{Narayan and Fisher}(1992)}]{narayan92}
\bibinfo{author}{\bibfnamefont{O.}~\bibnamefont{Narayan}} \bibnamefont{and}
  \bibinfo{author}{\bibfnamefont{D.~S.} \bibnamefont{Fisher}},
  \bibinfo{journal}{Phys. Rev. B} \textbf{\bibinfo{volume}{46}},
  \bibinfo{pages}{11520} (\bibinfo{year}{1992}).

\bibitem[{\citenamefont{Narayan and Fisher}(1993)}]{narayan93}
\bibinfo{author}{\bibfnamefont{O.}~\bibnamefont{Narayan}} \bibnamefont{and}
  \bibinfo{author}{\bibfnamefont{D.~S.} \bibnamefont{Fisher}},
  \bibinfo{journal}{Phys. Rev. B} \textbf{\bibinfo{volume}{48}},
  \bibinfo{pages}{7030} (\bibinfo{year}{1993}).

\bibitem[{\citenamefont{Narayan and Middleton}(1994)}]{narayan94}
\bibinfo{author}{\bibfnamefont{O.}~\bibnamefont{Narayan}} \bibnamefont{and}
  \bibinfo{author}{\bibfnamefont{A.~A.} \bibnamefont{Middleton}},
  \bibinfo{journal}{Phys. Rev. B} \textbf{\bibinfo{volume}{49}},
  \bibinfo{pages}{244} (\bibinfo{year}{1994}).

\bibitem[{\citenamefont{Chauve et~al.}(2000)\citenamefont{Chauve, Giamarchi,
  and Le~Doussal}}]{chauve00}
\bibinfo{author}{\bibfnamefont{P.}~\bibnamefont{Chauve}},
  \bibinfo{author}{\bibfnamefont{T.}~\bibnamefont{Giamarchi}},
  \bibnamefont{and}
  \bibinfo{author}{\bibfnamefont{P.}~\bibnamefont{Le~Doussal}},
  \bibinfo{journal}{Phys. Rev. B} \textbf{\bibinfo{volume}{62}},
  \bibinfo{pages}{6241} (\bibinfo{year}{2000}).

\bibitem[{\citenamefont{Chauve et~al.}(2001)\citenamefont{Chauve, Doussal, and
  Wiese}}]{chauve01}
\bibinfo{author}{\bibfnamefont{P.}~\bibnamefont{Chauve}},
  \bibinfo{author}{\bibfnamefont{P.~L.} \bibnamefont{Doussal}},
  \bibnamefont{and} \bibinfo{author}{\bibfnamefont{K.~J.} \bibnamefont{Wiese}},
  \bibinfo{journal}{Phys. Rev. Lett.} \textbf{\bibinfo{volume}{86}},
  \bibinfo{pages}{1785} (\bibinfo{year}{2001}).

\bibitem[{\citenamefont{Doussal et~al.}(2002)\citenamefont{Doussal, Wiese, and
  Chauve}}]{ledoussal02}
\bibinfo{author}{\bibfnamefont{P.~L.} \bibnamefont{Doussal}},
  \bibinfo{author}{\bibfnamefont{K.}~\bibnamefont{Wiese}}, \bibnamefont{and}
  \bibinfo{author}{\bibfnamefont{P.}~\bibnamefont{Chauve}},
  \bibinfo{journal}{Phys. Rev. B} \textbf{\bibinfo{volume}{66}},
  \bibinfo{pages}{174201} (\bibinfo{year}{2002}).

\bibitem[{\citenamefont{Santucci et~al.}(2010)\citenamefont{Santucci, Grob,
  Hansen, Toussaint, Schmittbuhl, and M{\aa}l{\o}y}}]{SantucciGHTSM14}
\bibinfo{author}{\bibfnamefont{S.}~\bibnamefont{Santucci}},
  \bibinfo{author}{\bibfnamefont{M.}~\bibnamefont{Grob}},
  \bibinfo{author}{\bibfnamefont{A.}~\bibnamefont{Hansen}},
  \bibinfo{author}{\bibfnamefont{R.}~\bibnamefont{Toussaint}},
  \bibinfo{author}{\bibfnamefont{J.}~\bibnamefont{Schmittbuhl}},
  \bibnamefont{and} \bibinfo{author}{\bibfnamefont{K.~J.}
  \bibnamefont{M{\aa}l{\o}y}}, \bibinfo{journal}{EPL (Europhysics Letters)}
  \textbf{\bibinfo{volume}{92}}, \bibinfo{pages}{44001} (\bibinfo{year}{2010}),
  \urlprefix\url{http://stacks.iop.org/0295-5075/92/i=4/a=44001}.

\bibitem[{\citenamefont{Rubio et~al.}(1989)\citenamefont{Rubio, Edwards,
  Dougherty, and Gollub}}]{RubioPRL1989}
\bibinfo{author}{\bibfnamefont{M.~A.} \bibnamefont{Rubio}},
  \bibinfo{author}{\bibfnamefont{C.~A.} \bibnamefont{Edwards}},
  \bibinfo{author}{\bibfnamefont{A.}~\bibnamefont{Dougherty}},
  \bibnamefont{and} \bibinfo{author}{\bibfnamefont{J.~P.}
  \bibnamefont{Gollub}}, \bibinfo{journal}{Phys. Rev. Lett.}
  \textbf{\bibinfo{volume}{63}}, \bibinfo{pages}{1685} (\bibinfo{year}{1989}),
  \urlprefix\url{http://link.aps.org/doi/10.1103/PhysRevLett.63.1685}.

\bibitem[{\citenamefont{Horvath et~al.}(1991)\citenamefont{Horvath, Family, and
  Vicsek}}]{HorvathJPhysA1991}
\bibinfo{author}{\bibfnamefont{V.~K.} \bibnamefont{Horvath}},
  \bibinfo{author}{\bibfnamefont{F.}~\bibnamefont{Family}}, \bibnamefont{and}
  \bibinfo{author}{\bibfnamefont{T.}~\bibnamefont{Vicsek}},
  \bibinfo{journal}{Journal of Physics A: Mathematical and General}
  \textbf{\bibinfo{volume}{24}}, \bibinfo{pages}{L25} (\bibinfo{year}{1991}),
  \urlprefix\url{http://stacks.iop.org/0305-4470/24/i=1/a=006}.

\bibitem[{\citenamefont{He et~al.}(1992)\citenamefont{He, Kahanda, and
  Wong}}]{hePRL1992}
\bibinfo{author}{\bibfnamefont{S.}~\bibnamefont{He}},
  \bibinfo{author}{\bibfnamefont{G.~L.} \bibnamefont{Kahanda}},
  \bibnamefont{and} \bibinfo{author}{\bibfnamefont{P.-z.} \bibnamefont{Wong}},
  \bibinfo{journal}{Phys. Rev. Lett.} \textbf{\bibinfo{volume}{69}},
  \bibinfo{pages}{3731} (\bibinfo{year}{1992}).

\bibitem[{\citenamefont{Buldyrev et~al.}(1992)\citenamefont{Buldyrev,
  Barab\'asi, Caserta, Havlin, Stanley, and Vicsek}}]{Buldyrev1992PRA}
\bibinfo{author}{\bibfnamefont{S.~V.} \bibnamefont{Buldyrev}},
  \bibinfo{author}{\bibfnamefont{A.-L.} \bibnamefont{Barab\'asi}},
  \bibinfo{author}{\bibfnamefont{F.}~\bibnamefont{Caserta}},
  \bibinfo{author}{\bibfnamefont{S.}~\bibnamefont{Havlin}},
  \bibinfo{author}{\bibfnamefont{H.~E.} \bibnamefont{Stanley}},
  \bibnamefont{and} \bibinfo{author}{\bibfnamefont{T.}~\bibnamefont{Vicsek}},
  \bibinfo{journal}{Phys. Rev. A} \textbf{\bibinfo{volume}{45}},
  \bibinfo{pages}{R8313} (\bibinfo{year}{1992}).

\bibitem[{\citenamefont{Family et~al.}(1992)\citenamefont{Family, Chan, and
  Amar}}]{Family1992}
\bibinfo{author}{\bibfnamefont{F.}~\bibnamefont{Family}},
  \bibinfo{author}{\bibfnamefont{K.~C.~B.} \bibnamefont{Chan}},
  \bibnamefont{and} \bibinfo{author}{\bibfnamefont{J.~G.} \bibnamefont{Amar}},
  in \emph{\bibinfo{booktitle}{Surface Disordering: Growth, Roughening and
  Phase Transitions}} (\bibinfo{publisher}{Nova Science}, \bibinfo{address}{New
  York}, \bibinfo{year}{1992}), pp. \bibinfo{pages}{205--212}.

\bibitem[{\citenamefont{Vicsek et~al.}(1990)\citenamefont{Vicsek, Cserzo, and
  Horvath}}]{vicsek1990physicaA}
\bibinfo{author}{\bibfnamefont{T.}~\bibnamefont{Vicsek}},
  \bibinfo{author}{\bibfnamefont{M.}~\bibnamefont{Cserzo}}, \bibnamefont{and}
  \bibinfo{author}{\bibfnamefont{V.~K.} \bibnamefont{Horvath}},
  \bibinfo{journal}{Physica A: Statistical Mechanics and its Applications}
  \textbf{\bibinfo{volume}{167}}, \bibinfo{pages}{315} (\bibinfo{year}{1990}).

\bibitem[{\citenamefont{Zhang et~al.}(1992)\citenamefont{Zhang, Zhang, Alstrom,
  and Levinsen}}]{zhang1992physicaA}
\bibinfo{author}{\bibfnamefont{J.}~\bibnamefont{Zhang}},
  \bibinfo{author}{\bibfnamefont{Y.-C.} \bibnamefont{Zhang}},
  \bibinfo{author}{\bibfnamefont{P.}~\bibnamefont{Alstrom}}, \bibnamefont{and}
  \bibinfo{author}{\bibfnamefont{M.~T.} \bibnamefont{Levinsen}},
  \bibinfo{journal}{Physica A: Statistical Mechanics and its Applications}
  \textbf{\bibinfo{volume}{189}}, \bibinfo{pages}{383} (\bibinfo{year}{1992}).

\bibitem[{\citenamefont{Le~Doussal and Wiese}(2003)}]{anisotropicFRG}
\bibinfo{author}{\bibfnamefont{P.}~\bibnamefont{Le~Doussal}} \bibnamefont{and}
  \bibinfo{author}{\bibfnamefont{K.~J.} \bibnamefont{Wiese}},
  \bibinfo{journal}{Phys. Rev. E} \textbf{\bibinfo{volume}{67}},
  \bibinfo{pages}{016121} (\bibinfo{year}{2003}).

\bibitem[{\citenamefont{Kim et~al.}(2003)\citenamefont{Kim, Choe, and
  Shin}}]{KimPRL03}
\bibinfo{author}{\bibfnamefont{D.-H.} \bibnamefont{Kim}},
  \bibinfo{author}{\bibfnamefont{S.-B.} \bibnamefont{Choe}}, \bibnamefont{and}
  \bibinfo{author}{\bibfnamefont{S.-C.} \bibnamefont{Shin}},
  \bibinfo{journal}{Phys. Rev. Lett.} \textbf{\bibinfo{volume}{90}},
  \bibinfo{pages}{087203} (\bibinfo{year}{2003}).

\bibitem[{\citenamefont{Ryu et~al.}(2007)\citenamefont{Ryu, Akinaga, and
  Shin}}]{ryu07nature}
\bibinfo{author}{\bibfnamefont{K.-S.} \bibnamefont{Ryu}},
  \bibinfo{author}{\bibfnamefont{H.}~\bibnamefont{Akinaga}}, \bibnamefont{and}
  \bibinfo{author}{\bibfnamefont{S.-C.} \bibnamefont{Shin}},
  \bibinfo{journal}{Nature Physics} \textbf{\bibinfo{volume}{3}},
  \bibinfo{pages}{547} (\bibinfo{year}{2007}).

\bibitem[{\citenamefont{Sethna}(2007)}]{Sethna07Crossover}
\bibinfo{author}{\bibfnamefont{J.~P.} \bibnamefont{Sethna}},
  \bibinfo{journal}{Nature Physics (News and Views)}
  \textbf{\bibinfo{volume}{3}}, \bibinfo{pages}{518} (\bibinfo{year}{2007}).

\bibitem[{\citenamefont{Bustingorry et~al.}(2010)\citenamefont{Bustingorry,
  Kolton, and Giamarchi}}]{BustingorryKG10}
\bibinfo{author}{\bibfnamefont{S.}~\bibnamefont{Bustingorry}},
  \bibinfo{author}{\bibfnamefont{A.~B.} \bibnamefont{Kolton}},
  \bibnamefont{and}
  \bibinfo{author}{\bibfnamefont{T.}~\bibnamefont{Giamarchi}},
  \bibinfo{journal}{Phys. Rev. B} \textbf{\bibinfo{volume}{82}},
  \bibinfo{pages}{094202} (\bibinfo{year}{2010}),
  \urlprefix\url{http://link.aps.org/doi/10.1103/PhysRevB.82.094202}.

\bibitem[{\citenamefont{Schmittbuhl et~al.}(2003)\citenamefont{Schmittbuhl,
  Hansen, and Batrouni}}]{SchmittbuhlCoalesce}
\bibinfo{author}{\bibfnamefont{J.}~\bibnamefont{Schmittbuhl}},
  \bibinfo{author}{\bibfnamefont{A.}~\bibnamefont{Hansen}}, \bibnamefont{and}
  \bibinfo{author}{\bibfnamefont{G.~G.} \bibnamefont{Batrouni}},
  \bibinfo{journal}{Phys. Rev. Lett.} \textbf{\bibinfo{volume}{90}},
  \bibinfo{pages}{045505} (\bibinfo{year}{2003}),
  \urlprefix\url{http://link.aps.org/doi/10.1103/PhysRevLett.90.045505}.

\bibitem[{\citenamefont{Laurson et~al.}(2010)\citenamefont{Laurson, Santucci,
  and Zapperi}}]{laurson2010}
\bibinfo{author}{\bibfnamefont{L.}~\bibnamefont{Laurson}},
  \bibinfo{author}{\bibfnamefont{S.}~\bibnamefont{Santucci}}, \bibnamefont{and}
  \bibinfo{author}{\bibfnamefont{S.}~\bibnamefont{Zapperi}},
  \bibinfo{journal}{Phys Rev E Stat Nonlin Soft Matter Phys}
  \textbf{\bibinfo{volume}{81}}, \bibinfo{pages}{046116}
  (\bibinfo{year}{2010}).

\bibitem[{\citenamefont{Schmittbuhl et~al.}(1995)\citenamefont{Schmittbuhl,
  Roux, Vilotte, and Jorgen~M\aa{}l\o{}y}}]{SchmittbuhlFlucLine}
\bibinfo{author}{\bibfnamefont{J.}~\bibnamefont{Schmittbuhl}},
  \bibinfo{author}{\bibfnamefont{S.}~\bibnamefont{Roux}},
  \bibinfo{author}{\bibfnamefont{J.-P.} \bibnamefont{Vilotte}},
  \bibnamefont{and}
  \bibinfo{author}{\bibfnamefont{K.}~\bibnamefont{Jorgen~M\aa{}l\o{}y}},
  \bibinfo{journal}{Phys. Rev. Lett.} \textbf{\bibinfo{volume}{74}},
  \bibinfo{pages}{1787} (\bibinfo{year}{1995}),
  \urlprefix\url{http://link.aps.org/doi/10.1103/PhysRevLett.74.1787}.

\bibitem[{\citenamefont{Rosso and Krauth}(2002{\natexlab{b}})}]{RussoK02}
\bibinfo{author}{\bibfnamefont{A.}~\bibnamefont{Rosso}} \bibnamefont{and}
  \bibinfo{author}{\bibfnamefont{W.}~\bibnamefont{Krauth}},
  \bibinfo{journal}{Phys. Rev. E} \textbf{\bibinfo{volume}{65}},
  \bibinfo{pages}{025101} (\bibinfo{year}{2002}{\natexlab{b}}),
  \urlprefix\url{http://link.aps.org/doi/10.1103/PhysRevE.65.025101}.

\bibitem[{\citenamefont{Nelson}(1975)}]{Nelson75}
\bibinfo{author}{\bibfnamefont{D.~R.} \bibnamefont{Nelson}},
  \bibinfo{journal}{Phys. Rev. B} \textbf{\bibinfo{volume}{11}},
  \bibinfo{pages}{3504} (\bibinfo{year}{1975}),
  \urlprefix\url{http://link.aps.org/doi/10.1103/PhysRevB.11.3504}.

\bibitem[{\citenamefont{Cardy}(1996)}]{Cardy96}
\bibinfo{author}{\bibfnamefont{J.}~\bibnamefont{Cardy}},
  \emph{\bibinfo{title}{Scaling and renormalization in statistical physics}},
  vol.~\bibinfo{volume}{5} (\bibinfo{publisher}{Cambridge University Press},
  \bibinfo{year}{1996}).

\bibitem[{\citenamefont{Amaral et~al.}(1995)\citenamefont{Amaral, Barab\'asi,
  Makse, and Stanley}}]{amaral95}
\bibinfo{author}{\bibfnamefont{L.~A.~N.} \bibnamefont{Amaral}},
  \bibinfo{author}{\bibfnamefont{A.-L.} \bibnamefont{Barab\'asi}},
  \bibinfo{author}{\bibfnamefont{H.~A.} \bibnamefont{Makse}}, \bibnamefont{and}
  \bibinfo{author}{\bibfnamefont{H.~E.} \bibnamefont{Stanley}},
  \bibinfo{journal}{Phys. Rev. E} \textbf{\bibinfo{volume}{52}},
  \bibinfo{pages}{4087} (\bibinfo{year}{1995}),
  \urlprefix\url{http://link.aps.org/doi/10.1103/PhysRevE.52.4087}.

\bibitem[{\citenamefont{Durin and Zapperi}(2006)}]{DurinZ06}
\bibinfo{author}{\bibfnamefont{G.}~\bibnamefont{Durin}} \bibnamefont{and}
  \bibinfo{author}{\bibfnamefont{S.}~\bibnamefont{Zapperi}}, in
  \emph{\bibinfo{booktitle}{The Science of Hysteresis, Vol. II}}
  (\bibinfo{publisher}{Academic Press}, \bibinfo{year}{2006}), pp.
  \bibinfo{pages}{181--267}.

\bibitem[{\citenamefont{Narayan}(1996)}]{narayan96}
\bibinfo{author}{\bibfnamefont{O.}~\bibnamefont{Narayan}},
  \bibinfo{journal}{Phys. Rev. Lett.} \textbf{\bibinfo{volume}{77}},
  \bibinfo{pages}{3855} (\bibinfo{year}{1996}),
  \urlprefix\url{http://link.aps.org/doi/10.1103/PhysRevLett.77.3855}.

\bibitem[{\citenamefont{Chen et~al.}(2011)\citenamefont{Chen, Papanikolaou,
  Sethna, Zapperi, and Durin}}]{ChenPSZD11}
\bibinfo{author}{\bibfnamefont{Y.}~\bibnamefont{Chen}},
  \bibinfo{author}{\bibfnamefont{S.}~\bibnamefont{Papanikolaou}},
  \bibinfo{author}{\bibfnamefont{J.~P.} \bibnamefont{Sethna}},
  \bibinfo{author}{\bibfnamefont{S.}~\bibnamefont{Zapperi}}, \bibnamefont{and}
  \bibinfo{author}{\bibfnamefont{G.}~\bibnamefont{Durin}},
  \bibinfo{journal}{Physical Review E} \textbf{\bibinfo{volume}{84}},
  \bibinfo{pages}{061103} (\bibinfo{year}{2011}).

\bibitem[{\citenamefont{Jost and Usadel}(1998)}]{JostautomatonqEW97}
\bibinfo{author}{\bibfnamefont{M.}~\bibnamefont{Jost}} \bibnamefont{and}
  \bibinfo{author}{\bibfnamefont{K.}~\bibnamefont{Usadel}},
  \bibinfo{journal}{Physica A: Statistical Mechanics and its Applications}
  \textbf{\bibinfo{volume}{255}}, \bibinfo{pages}{15 } (\bibinfo{year}{1998}),
  ISSN \bibinfo{issn}{0378-4371},
  \urlprefix\url{http://www.sciencedirect.com/science/article/pii/S0378437198001125}.

\bibitem[{\citenamefont{Rosso et~al.}(2007)\citenamefont{Rosso, Le~Doussal, and
  Wiese}}]{FRGDepinningRosso}
\bibinfo{author}{\bibfnamefont{A.}~\bibnamefont{Rosso}},
  \bibinfo{author}{\bibfnamefont{P.}~\bibnamefont{Le~Doussal}},
  \bibnamefont{and} \bibinfo{author}{\bibfnamefont{K.~J.} \bibnamefont{Wiese}},
  \bibinfo{journal}{Phys. Rev. B} \textbf{\bibinfo{volume}{75}},
  \bibinfo{pages}{220201} (\bibinfo{year}{2007}),
  \urlprefix\url{http://link.aps.org/doi/10.1103/PhysRevB.75.220201}.

\bibitem[{\citenamefont{Lee and Kim}(2005)}]{LeeK05}
\bibinfo{author}{\bibfnamefont{C.}~\bibnamefont{Lee}} \bibnamefont{and}
  \bibinfo{author}{\bibfnamefont{J.~M.} \bibnamefont{Kim}},
  \bibinfo{journal}{J. Korean Phys. Soc.} \textbf{\bibinfo{volume}{47}},
  \bibinfo{pages}{13} (\bibinfo{year}{2005}).

\bibitem[{\citenamefont{Leschhorn and Tang}(1993)}]{PRLZetaComment}
\bibinfo{author}{\bibfnamefont{H.}~\bibnamefont{Leschhorn}} \bibnamefont{and}
  \bibinfo{author}{\bibfnamefont{L.-H.} \bibnamefont{Tang}},
  \bibinfo{journal}{Phys. Rev. Lett.} \textbf{\bibinfo{volume}{70}},
  \bibinfo{pages}{2973} (\bibinfo{year}{1993}),
  \urlprefix\url{http://link.aps.org/doi/10.1103/PhysRevLett.70.2973}.

\bibitem[{\citenamefont{L\'opez et~al.}(1997)\citenamefont{L\'opez, Rodriguez,
  and Cuerno}}]{LopezRC97}
\bibinfo{author}{\bibfnamefont{J.~M.} \bibnamefont{L\'opez}},
  \bibinfo{author}{\bibfnamefont{M.~A.} \bibnamefont{Rodriguez}},
  \bibnamefont{and} \bibinfo{author}{\bibfnamefont{R.}~\bibnamefont{Cuerno}},
  \bibinfo{journal}{Phys. Rev. E} \textbf{\bibinfo{volume}{56}},
  \bibinfo{pages}{3993} (\bibinfo{year}{1997}),
  \urlprefix\url{http://link.aps.org/doi/10.1103/PhysRevE.56.3993}.

\bibitem[{\citenamefont{L\'opez and Schmittbuhl}(1998)}]{LopezS98}
\bibinfo{author}{\bibfnamefont{J.~M.} \bibnamefont{L\'opez}} \bibnamefont{and}
  \bibinfo{author}{\bibfnamefont{J.}~\bibnamefont{Schmittbuhl}},
  \bibinfo{journal}{Phys. Rev. E} \textbf{\bibinfo{volume}{57}},
  \bibinfo{pages}{6405} (\bibinfo{year}{1998}),
  \urlprefix\url{http://link.aps.org/doi/10.1103/PhysRevE.57.6405}.

\bibitem[{\citenamefont{Ramasco et~al.}(2000)\citenamefont{Ramasco, L\'opez,
  and Rodriguez}}]{Ramasco00}
\bibinfo{author}{\bibfnamefont{J.~J.} \bibnamefont{Ramasco}},
  \bibinfo{author}{\bibfnamefont{J.~M.} \bibnamefont{L\'opez}},
  \bibnamefont{and} \bibinfo{author}{\bibfnamefont{M.~A.}
  \bibnamefont{Rodriguez}}, \bibinfo{journal}{Phys. Rev. Lett.}
  \textbf{\bibinfo{volume}{84}}, \bibinfo{pages}{2199} (\bibinfo{year}{2000}),
  \urlprefix\url{http://link.aps.org/doi/10.1103/PhysRevLett.84.2199}.

\bibitem[{\citenamefont{Frederiksen et~al.}(2004)\citenamefont{Frederiksen,
  Jacobsen, Brown, and Sethna}}]{FrederiksenJBS04}
\bibinfo{author}{\bibfnamefont{S.~L.} \bibnamefont{Frederiksen}},
  \bibinfo{author}{\bibfnamefont{K.~W.} \bibnamefont{Jacobsen}},
  \bibinfo{author}{\bibfnamefont{K.~S.} \bibnamefont{Brown}}, \bibnamefont{and}
  \bibinfo{author}{\bibfnamefont{J.~P.} \bibnamefont{Sethna}},
  \bibinfo{journal}{Physical Review Letters} \textbf{\bibinfo{volume}{93}},
  \bibinfo{pages}{216401} (\bibinfo{year}{2004}).

\end{thebibliography}

\end{document}